%% file: ms.tex
\documentclass[12pt,preprint]{aastex}
\begin{document}
\newcommand{\msun}{\mbox{M$_{\odot}$}}
\newcommand{\rsun}{\mbox{R$_{\odot}$}}
\newcommand{\lsun}{\mbox{L$_{\odot}$}}
\title{The First DIRECT Distance Determination to a Detached Eclipsing
Binary in M33\altaffilmark{1}}

\author{A. Z. Bonanos\altaffilmark{2}, K. Z. Stanek\altaffilmark{3},
R. P. Kudritzki\altaffilmark{4}, L.M. Macri\altaffilmark{5,6},
D. D. Sasselov\altaffilmark{7}, \\ J. Kaluzny\altaffilmark{8},
P. B. Stetson\altaffilmark{9}, D. Bersier\altaffilmark{10},
F. Bresolin\altaffilmark{4}, T. Matheson\altaffilmark{5},\\
B.J. Mochejska\altaffilmark{8}, N. Przybilla\altaffilmark{4,11},
A.H. Szentgyorgyi\altaffilmark{7}, J. Tonry\altaffilmark{4},
G. Torres\altaffilmark{7}}

\altaffiltext{1}{Based on observations obtained with 1.2 meter FLWO,
2.1 meter KPNO, 3.5 meter WIYN, 8.2 meter Gemini, and 10 meter Keck-2
telescopes.}
\altaffiltext{2}{Vera Rubin Fellow, Carnegie Institution of Washington,
Department of Terrestrial Magnetism, 5241 Broad Branch Road NW,
Washington, DC 20015; bonanos@dtm.ciw.edu}
\altaffiltext{3}{Department of Astronomy, The Ohio State University, 140
W. 18th Avenue, Columbus, OH 43210; kstanek@astronomy.ohio-state.edu }
\altaffiltext{4}{Institute for Astronomy, University of Hawaii, 2680
Woodlawn Drive, Honolulu, HI~96822; kud, bresolin, jt@ifa.hawaii.edu.}
\altaffiltext{5}{National Optical Astronomy Observatory, 950 North
Cherry Avenue, Tucson, AZ 85719; lmacri, matheson@noao.edu.}
\altaffiltext{6}{Hubble Fellow \& Goldberg Fellow.}
\altaffiltext{7}{Harvard-Smithsonian Center for Astrophysics, 60 Garden
St., Cambridge, MA~02138; sasselov, aszentgyorgyi,
torres@cfa.harvard.edu.}
\altaffiltext{8}{Copernicus Astronomical Center, Bartycka 18, 00-716
Warszawa, Poland; jka, mochejsk@camk.edu.pl.}
\altaffiltext{9}{National Research Council of Canada, Herzberg Institute
of Astrophysics, 5071 West Saanich Road, Victoria, BC V9E 2E7,
Canada; peter.stetson@nrc-cnrc.gc.ca}
\altaffiltext{10}{Liverpool John Moores University, Twelve Quays House, 
Egerton Wharf, Birkenhead, CH41 1LD, United Kingdom; dbf@astro.livjm.ac.uk.}
\altaffiltext{11}{Dr. Remeis-Sternwarte, Astronomical Insitute of the
University of Erlangen-Nuremberg, Sternwartstrasse 7, D-96049 Bamberg,
Germany; przybilla@sternwarte.uni-erlangen.de.}

\begin{abstract}

We present the first direct distance determination to a detached
eclipsing binary in M33, which was found by the DIRECT Project. Located
in the OB~66 association at coordinates ($\alpha,
\delta)=(01\!\!:\!\!33\!\!:\!\!46.17, +30\!\!:\!\!44\!\!:\!\!39.9$) for
J2000.0, it was one of the most suitable detached eclipsing binaries
found by DIRECT for distance determination, given its apparent magnitude
and orbital period. We obtained follow-up $BV$ time series photometry,
$JHK_s$ photometry and optical spectroscopy from which we determined the
parameters of the system. It contains two O7 main sequence stars with
masses of $33.4\pm3.5 \; \msun$ and $30.0\pm3.3 \; \msun$ and radii of
$12.3\pm0.4\; \rsun$ and $8.8\pm0.3\; \rsun$, respectively. We derive
temperatures of $37000\pm1500$ K and $35600\pm1500$ K. Using $BVRJHK_s$
photometry for the flux calibration, we obtain a distance modulus of
$24.92\pm0.12$ mag ($964\pm54$ kpc), which is $\sim$0.3 mag longer than
the Key Project distance to M33. We discuss the implications of our
result and the importance of establishing M33 as an independent rung on
the cosmological distance ladder.

\end{abstract}


\section{Introduction}

Starting in 1996 we undertook a long term project, DIRECT
(i.e. ``direct distances''), to obtain the distances to two important
galaxies in the cosmological distance ladder, M31 and M33. These
``direct'' distances are obtained by measuring the absolute distance
to detached eclipsing binaries (DEBs).

M31 and M33 are the nearest and most suitable Local Group galaxies for
calibrating the extragalactic distance scale. However, they present a
much greater observational challenge than the current anchor of the
distance scale, the Large Magellanic Cloud (LMC). Their greater distance
makes the brightest stars in them appear $\sim 6$ mag fainter than the
brightest LMC stars, thus pushing the limit of current spectroscopic
capabilities. In addition, crowding and blending become more significant
with increasing distance \citep{Mochejska00,
Mochejska01c}. Unfortunately, distances are now known to no better than
10-15\%, as there are discrepancies of $0.2-0.3\;{\rm mag}$ between
various distance indicators \citep[e.g.][Figure 8]{Benedict02}. These
uncertainties limit the calibration of stellar luminosities and
population synthesis models for early galaxy formation and evolution.

DEBs have the potential to establish distances to M31 and M33 with an
unprecedented accuracy of 5\% \citep[for reviews and history of method
see][]{Andersen91, Hilditch96, Paczynski97, Kruszewski99}. They offer a
single step distance determination to nearby galaxies and may therefore
provide an accurate zero point calibration of various distance
indicators -- a major step towards very accurate and independent
determination of the Hubble constant. In the last few years, eclipsing
binaries have been used to obtain accurate distance estimates to the
Large Magellanic Cloud \citep[e.g.][]{Guinan98,Fitzpatrick03}, the Small
Magellanic Cloud \citep{Harries03,Hilditch05} and most recently to a
semi-detached system in M31 \citep{Ribas05}. Distances to individual
eclipsing binaries in the Magellanic Clouds are claimed to be accurate
to better than $5\%$.

Detached eclipsing binaries have yet to be used as distance indicators
to M31 and M33. The DIRECT project has initiated a search for DEBs and
new Cepheids in the M31 and M33 galaxies. We have analyzed five
$11\arcmin\times11\arcmin$ fields in M31, A-D and F \citep[][hereafter
Papers I, II, III, IV, V]{Kaluzny98, Stanek98, Stanek99, Kaluzny99,
Mochejska99} and one $22\arcmin\times22\arcmin$ field, Y
\citep[][hereafter Paper IX]{Bonanos03}. A total of 674 variables,
mostly new, were found: 89 eclipsing binaries, 332 Cepheids and 253
other periodic, possible long-period or non-periodic variables. We
have analyzed two fields in M33, A and B \citep[][hereafter Paper
VI]{Macri01b} and found 544 variables: 47 eclipsing binaries, 251
Cepheids and 246 other variables. Follow up observations of fields
M33A and M33B produced 280 and 612 new variables, respectively
\citep[][hereafter Papers VII, VIII]{Mochejska01a, Mochejska01b},
including 101 new eclipsing binaries. Variables from two more DIRECT
fields, one in M31 and the other in M33, remain to be reported.

Of the 237 eclipsing binaries found by DIRECT, only four are bright
enough ($V_{max}<20$ mag) for distance determination with currently
available telescopes. An additional criterion for selection is that they
contain deep eclipses, which removes degeneracies in the modeling.
D33J013346.2+304439.9 is the brightest of these, discovered in field
M33A (Paper VI), and this paper presents the distance we obtained to it
with subsequent observations. In \S 2 we describe the observations and
the data reduction, in \S 3 we present the light curve and radial
velocity curve analysis, in \S 4 the distance determination and in \S 5
the discussion.

\section{Observations}

The DIRECT Project discovered the detached eclipsing binary
D33J013346.2+304439.9 ($\alpha=01$:33:46.17, $\delta= +30$:44:39.9:
J2000.0) in field M33A (Paper VI), using the F.L. Whipple Observatory
1.2 meter telescope between 1996 September and 1997 October.

\subsection{Follow-up $V$ and $B$-band Photometry}

In 1999 and 2001 we obtained follow-up photometry with the Tektronix
$2048\times2048$ CCD (T2KA camera) at the KPNO 2.1 m telescope, with a
pixel scale of $0\farcs305\; \rm pixel^{-1}$. The 1999 observations are
described in Paper VII. We additionally obtained $94\times600s$
exposures in the $V$-band and $19\times600s$ in $B$-band in October
2001. Tables~\ref{Vlightcurve} and \ref{Blightcurve} present the $BV$
light curves. The images were processed with standard IRAF\footnote{IRAF
is distributed by the National Optical Astronomy Observatory, which are
operated by the Association of Universities for Research in Astronomy,
Inc., under cooperative agreement with the NSF.}  routines. The
nonlinearity of the CCD was corrected with the method outlined in Paper
VII.

The photometry for the variable stars was extracted using the ISIS
image subtraction package \citep{Alard98,Alard00} from the $V$ and
$B$-band data. We followed the ISIS reduction procedure described in
detail in Paper VII. DAOPHOT/ALLSTAR PSF photometry was performed on
the $B$ and $V$ reference image separately and aperture corrections
were applied before converting the flux light curves to instrumental
magnitudes. In addition, we performed DAOPHOT PSF photometry on all
frames to verify that the shapes of the light curves were correctly
measured with ISIS.

Observations of the DEB were also obtained with the MiniMosaic CCD
camera on the WIYN\footnote{The WIYN Observatory is a joint facility of
the University of Wisconsin-Madison, Indiana University, Yale
University, and the National Optical Astronomy Observatory.} 3.5 meter
telescope at KPNO, with a pixel scale of $0\farcs28 \;\rm
pixel^{-1}$. On 1999 October 3, we obtained a $300s$ exposure in each of
the $V$ and $B$-bands at airmass 1.28 and 1.25, respectively and at
phase 0.73. Conditions were photometric, and 3 Landolt standard fields
were observed, covering a range in airmass from 1.17 to 1.40.  These
data were used for calibrating the $BV$ light curves to standard
photometric bands \citep{Landolt92}.

\subsection{Photometric Calibration}

The aperture photometry of the stars in the \citet{Landolt92} fields was
derived with DAOGROW \citep{Stetson90}. The photometric solutions were
very robust ($\rm rms\sim0.03$ mag). Instrumental magnitudes were
derived with DAOPHOT/ALLSTAR and aperture corrections with DAOGROW after
subtraction of all but PSF stars. As for standards the largest aperture
was set to 24 pixels. The WIYN calibration yielded $V=19.51\pm0.01$ mag,
$B-V=-0.20\pm0.01$ mag for the DEB.

Other surveys of M33 have measured photometry for the DEB: \citet[][Star
UIT196]{Massey96} and \citet[][Star 929]{Ivanov93}, however the quality
of the data is inferior. The Local Group Survey \citep{Massey06} has
obtained high quality photometry of the DEB, using the KPNO 4-meter
telescope. They measured out of eclipse magnitudes of $V=19.52\pm0.01$,
$B-V=-0.20\pm0.01$, $U-B=-1.14\pm0.01$, $V-R=-0.12\pm0.01$,
$R-I=-0.18\pm0.01$, in excellent agreement with our $BV$
values. Henceforth, we use the average of these values: $B=19.32\pm0.01$
and $V=19.52\pm0.01$.

\subsection{Near-Infrared Photometry}

We carried out near-infrared observations of the DEB using the Gemini
North telescope and NIRI \citep{Hodapp03} as queue program
GN-2005B-DD-4. We observed the system in the $J$, $H$ and $K_s$
bandpasses on 2006 January 03 (UT) between phases $0.39-0.40$.
Conditions were photometric and the range of airmasses was between 1.05
and 1.25. The typical PSF FWHM was $0\farcs3$, or 2.5 pix at the $f/6$
plate scale of NIRI. Total on-source exposure times were 11 minutes in
$J$, 35 minutes in $H$, and 32 minutes in $K_s$, with individual
exposures of 30$s$ in $J$ and $H$ and 60$s$ in $K_s$.

We reduced the images (bad pixel masking, dark current correction, sky
subtraction and flat fielding) using the IRAF Gemini NIRI package
(v1.8). We performed PSF photometry using DAOPHOT and ALLSTAR
\citep{Stetson87} on each individual image, because the PSF exhibited
significant variations in ellipticity from image to image. We defined
the PSF out to a radius of 8 pixels ($0\farcs94$) which included
essentially all of the stellar flux. Aperture corrections were derived
using DAOGROW \citep{Stetson90} and amounted to $\lesssim 0.02$~mag.

We calibrated the photometry using several 2MASS stars in the field of
view that were determined to be isolated from our higher-resolution
Gemini images. In some cases, we added the flux of faint neighboring
stars to that of the bright comparison star, since they appear
unresolved in the 2MASS images. We determined effective zeropoints for
this Gemini/NIRI dataset (at airmass $\sim 1.1$ and $J-K_{s}\sim 1$) of
23.82, 23.75, and 23.23$\pm0.05$~mag for $J $, $H$ and $K_s$,
respectively (corresponding to the 2MASS magnitude of a star that would
yield 1 ADU/sec). By following this calibration path, we found mean
out-of-eclipse magnitudes for the DEB of $J= 20.02\pm0.05$,
$H=20.05\pm0.06$ and $K_s=20.13\pm0.06$~mag.

\begin{figure}[ht]  
\includegraphics[angle=270,width=7in]{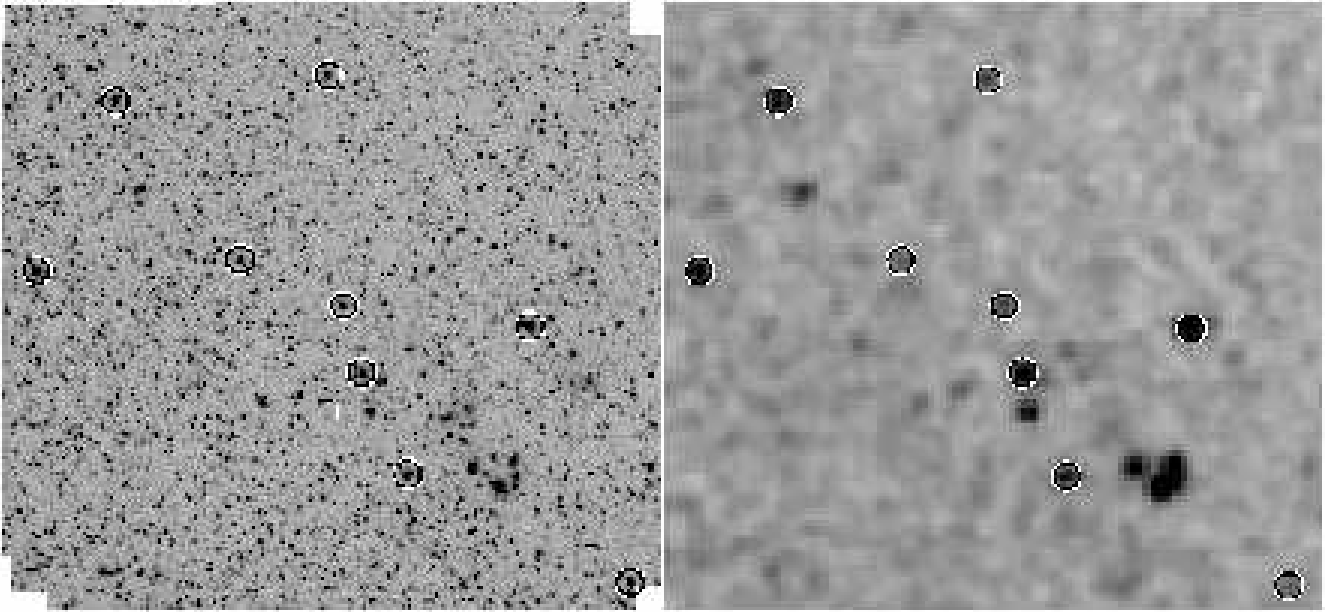}
\caption{$K_s-$band finder chart, $1\farcs8\times1\farcs8$, N is up, E
to the left. The circles indicate the stars used for calibration on both
the NIRI image (Left) and the 2MASS image (Right). The DEB is marked on
the NIRI image by the crosshairs.}
\label{deb_niri_img}
\end{figure}

Figure~\ref{deb_niri_img} is a $K_s-$band finder chart of the field,
indicating the location of the DEB and the photometric calibration
reference stars in both the Gemini/NIRI and 2MASS images.

\subsection{Spectroscopic Observations}

We obtained spectra of the DEB over 7 epochs, 6 half and 1 full night
(2004 Oct 11 UT), in the fall of 2002, 2003 and 2004 with the Echellette
Spectrograph and Imager \citep[or ESI,][]{Sheinis02} on the 10-meter
Keck-II telescope on Mauna Kea. In the fall of 2004, we also obtained 4
sets of $5\times 2700$ s spectra at quadrature with the B1200 lines/mm
grating of the Gemini Multi Object Spectrograph \citep[GMOS,][]{Hook03}
on the Gemini-North 8-meter telescope on Mauna
Kea. Table~\ref{debtab:obs} summarizes the exposure times and dates of
observations, which total 4 nights on Keck and 19 hours on Gemini. The
ESI spectra range from $ \rm 3950-10000 \AA$, have a reciprocal
dispersion of 11 $\rm km\; s^{-1}/pix$ and with a $1\arcsec$ slit we
achieved a velocity resolution of 74 $\rm km \; s^{-1}$. The GMOS
spectra range from $\rm 3770-5227\AA$, have a 0.23 $\rm \AA/pix$
dispersion and with a $0\farcs75$ slit we achieved a resolution of 120
$\rm km \; s^{-1}$. The signal to noise (S/N) of the spectra ranges from
15-40. Observations were taken at low airmass and the slit was rotated
75 deg East from the North to minimize contamination from nearby
\ion{H}{2} regions.

Standard CCD processing was done with IRAF. We used the algorithm of
\citet{Pych04} for cosmic ray rejection from each two dimensional
image. We then optimally extracted the spectra and calibrated them in
wavelength with IRAF routines. Three copper-argon calibration spectra
were interspersed between the 5 sets of GMOS spectra. In 2002, we found
ESI to be stable to a fraction of a pixel, thus in subsequent runs we
only took several CuAr and HgNeXe lamps at the beginning and end of the
night (or half night) that we averaged together and used to establish
the wavelength solution, with low-order polynomial fits to the
lamps. The small spatial extent of the ESI slit made background
subtraction difficult at the Balmer lines. Such contamination does not
affect the lines used for radial velocity determination,
however. Spectra of standard stars were obtained at each epoch with ESI
and were used to calibrate the DEB in flux and to remove telluric lines
with our IDL routines \citep{Wade88, Matheson00}.

We used the IRAF Gemini package v1.7 to reduce the GMOS data. We
wavelength calibrated the 3 extensions/chips of each exposure separately
with copper-argon lamps for a more accurate wavelength calibration and
merged the extensions at the end. The gaps between chips correspond to
the regions $4244-4255 \; \rm \AA$ and $4731-4747\; \rm \AA$ in the
spectrum. For the flux calibration we used standard star spectra that
were observed nearest in time to the DEB. The observations from each
night were averaged together, weighting by the exposure times. Finally,
we normalized the combined spectra, dividing them by a spline function
which we fit interactively to points in the continuum region.

Absorption lines from both stars are clearly resolved in the
spectrum. We were not able to avoid emission from nearby \ion{H}{2}
regions, thus our spectra contain strong nebular emission in the
hydrogen lines and weaker emission in some of the helium lines, as well
as forbidden oxygen lines [\ion{O}{3}] $\lambda\lambda$4959, 5007. The
GMOS spectra show broad \ion{Ca}{2} $\lambda$3933.7 absorption and a
narrow absorption line at $\lambda4433.3$, possibly due to an
interstellar band.

\section{Light Curve and Radial Velocity Curve Analysis}

Initially, we used the multiharmonic analysis of variance technique
\citep{Schwarzenberg89, Schwarzenberg96} to search the light curves for
periodicity. We merged the $B$ and $V$ light curves, scaled to $V$, and
searched for the best period by fitting a Gaussian to the peak of the
periodogram generated by this method and found a period $P=4.8935$ days,
which serves as the initial guess for the modeling. In this paper, we
define the primary star (star 1) photometrically, as the hotter star
producing the deeper eclipse at phase zero. In the analysis below, we
show the primary to be the larger and more massive star also.

\subsection{Initial Light Curve Parameter Estimation}

We first estimated the parameters of the $V$-band light curve with the
EBOP code \citep{Nelson72,Popper81}. We computed fits for several
periods and found the fit with a period of $P= 4.8938$~days to have
the smallest residuals. Bootstrap analysis \citep{Press92} yielded
very small values for the errors, $\sim 2\%$, which are smaller than
errors from the radial velocity curves discussed later on.

EBOP's biaxial ellipsoidal approximation holds for values of oblateness
$\epsilon$ less than 0.04 and fractional radii less than 0.2. We find
$\epsilon=0.02$, but a fractional radius for the primary slighty larger
than 0.2, thus we proceed to a more detailed treatment. Using the EBOP
parameters as starting values, we fit the 83 $B$-band points and the 278
$V$-band points in the light curve simultaneously with the April 2003
version of the Wilson-Devinney (WD) program \citep{Wilson71, Wilson79,
Wilson90, vanHamme03} in the detached mode (MODE=2). We found the
following best fit parameters: eccentricity $e=0.17$, $\omega=251.8$
deg, inclination $i=86.9$ deg, light ratio $L_{2}/L_{1}= 0.4912$ in $V$
and 0.4923 in $B$ and flux ratio $F_{1}/F_{2}=1.039$ in $V$ and 1.036 in
$B$, in agreement with the definition of the primary being the hotter
star. These preliminary values were refined later (see \S\ref{combwd})
in a simultaneous fit of both light curves and radial velocity
curves. In eccentric binaries, there is a degeneracy in determining the
radius and luminosity ratio from the light curves alone. We found that
reversing the radii so that the primary is smaller and less luminous
also produced a good light curve fit. We resolve this by using the
spectroscopic light ratio in \S\ref{fastwind}.

\subsection{TODCOR Analysis of Spectra}

We used the method of two dimensional cross correlation or TODCOR,
developed by \citet{Zucker94}, to measure radial velocities of the stars
in the DEB. TODCOR can distinguish small velocity separations even more
accurately than one dimensional cross correlation. Initially, we
calculated a grid of template spectra over a range of temperatures
around 30000 K using the LTE ATLAS9 models and opacities developed by
\citet{Kurucz93}. We ran TODCOR with ATLAS9 template spectra and found
preliminary values for the semi-amplitudes, mass ratio and thus the
semi-major axis. We thus obtained estimates for $\log(g)$ and the masses
from which we computed a grid of non-local thermodynamic equilibrium
(NLTE) spectra with FASTWIND \citep{Santolaya97, Puls05}. This
hydrodynamic NLTE code includes stellar winds, spherical extension and
metal line blanketing of millions of lines in (approximate) NLTE. We
found the best fit spectra to have effective temperatures T$_{\rm
eff1}=37000\pm1500$~K and T$_{\rm eff2}=35600\pm1500$~K. A detailed
description of the temperature determination follows in
\S\ref{fastwind}.

The NLTE spectra were used to derive the final radial velocities of the
stars. We prepared them for TODCOR by applying rotational broadening of
$\rm 120\; km\; s^{-1}$ which was determined from the spectral fits. We
also applied instrumental broadening to match the resolution of the
observations, by convolving the models with a Gaussian of appropriate
FWHM for ESI and GMOS. The IRAF $rvsao.xcsao$ \citep{Tonry79, Kurtz92}
task was used to compute the one dimensional cross correlation function,
required by TODCOR, for each observed spectrum with the appropriate
model. We used the range $\rm 3975-6000\;\AA$ for ESI spectra and $\rm
3950-5227\;\AA$ for GMOS spectra, excluding the Balmer lines and
[\ion{O}{3}] emission lines. We ran TODCOR with model spectra containing
only hydrogen and helium lines, masking out the broad hydrogen lines,
and assuming a luminosity ratio $\rm L_{2}/L_{1} = 0.49$ found from the
light curve analysis. Table~\ref{vel} presents the measured velocities
for each spectrum. The spectrum of 20021101 at phase 0.595 produced a
spurious velocity for the secondary star. We excluded it from further
analysis because this spectrum has the lowest $S/N$ ($\sim$ 15) and the
secondary contributes only half the amount of light to the spectrum
which makes it even harder to measure. The velocities measured are
robust and accurate to $\rm 30\;km\; s^{-1}$. As a final test we ran
TODCOR using ATLAS9 models (37000K, 36000K with $\log(g)=4$) and TLUSTY
\citep{Lanz03} models including metals (37500K for both stars and
$\log(g)=3.75$ and 4.00). Both runs produced radial velocities and
semi-amplitudes in agreement within errors.

From the radial velocity curve analysis alone, we found a mass ratio
$q=M_2/M_1=0.90\pm0.06$, velocity semi-amplitudes of $\rm K_{1}=240
\pm 11\; km \; s^{-1}$ and $\rm K_{2}=268 \pm 11\; km \; s^{-1}$,
systemic velocity $\gamma=-214.1\pm 6.8 \rm \; km \; s^{-1}$ and
semi-major axis $a \sin i=48.3 \pm 1.6 \; \rsun$ and minimum masses of
$M_{1} \sin^{3}i=33.3 \pm 3.5\; \msun$ and $M_{2}\sin^{3}i=29.9 \pm
3.3\; \msun$, fixing $e$ and $\omega$ from the preliminary light curve
analysis described above.

Measuring radial velocities of early-type stars is complex for several
reasons: they have very few lines and the strongest of these, mainly the
Balmer lines, are broadened due to rotation and the Stark effect. The
standard one dimensional cross correlation method has been shown to
produce systematically smaller semi-amplitudes when blended lines are
included. \citet{Hilditch01} estimates the effect to be $\sim50\%$ for
hydrogen lines and $\sim10\%$ for helium lines, for velocity differences
between components of less than $200\; \rm km \; s^{-1}$. However, with
TODCOR systematic errors due to blending are avoided. Finally, the
rotational velocity used in the synthetic spectrum is also a source of
systematic error, if it deviates from the true value.  The value of $v
\sin i$ fit to the spectra is accurate to $\sim \rm 20 \; km\; s^{-1}$.

We attempted to use the spectral disentangling technique \citep{Simon94}
which has been shown to be superior to standard one dimensional cross
correlation by \citet{Harmanec97} in yielding more accurate
velocities. We used the public code FDBinary \citep{Ilijic01, Ilijic04},
but found significantly different semi-amplitudes than those from the
TODCOR analysis with unrealistically small errors and concluded that
higher signal to noise spectra ($S/N\geq40$) at all epochs are required
for this method to give meaningful results. The presence of nebular
emission lines, even though masked out, may also have hindered this
method.

\subsection{Combined Analysis of LC and RV curves}
\label{combwd}

With the radial velocity measurements in hand, we proceeded to perform
a simultaneous fit of the $BV$ light curves and the radial velocity
curves with WD in the detached mode (MODE 2) for 13 parameters: the
inclination $i$, eccentricity $e$, the argument of periastron
$\omega_{0}$, the semi-major axis $a$, system center of mass radial
velocity $\gamma$, the surface potential for each star $\Omega_{1}$
and $\Omega_{2}$, mass ratio $q=M_{2}/M_{1}$, T$_{\rm eff2}$, period
$P$, time of primary eclipse T$_0$, and band pass luminosity $L_{1}$
in each band.

Following the advice of \citet{vanHamme93}, we used the square root limb
darkening law which gives better results for hot stars in the
optical. Theoretical bolometric and passband specific limb darkening
coefficients were taken from theoretical values from \citet{Claret00}
for a LTE ATLAS9 stellar atmosphere model with $\rm T_{\rm eff}=37000$
K, $\log(g)=4.0$ (cgs), turbulent velocity of 4 $\rm km \; s^{-1}$ and
solar metallicity. We fixed gravity brightening exponents to unity and
albedos to 0.5 from theoretical values \citep{Hilditch01} for stars at
such temperatures. For reflection, we used a MREF value of 1 for simple
treatment of reflection with the inverse square law. Additionally, we
allowed for non Keplerian effects on the radial velocity curve. All data
points were weighted equally.

\begin{figure}[ht]  
\plotone{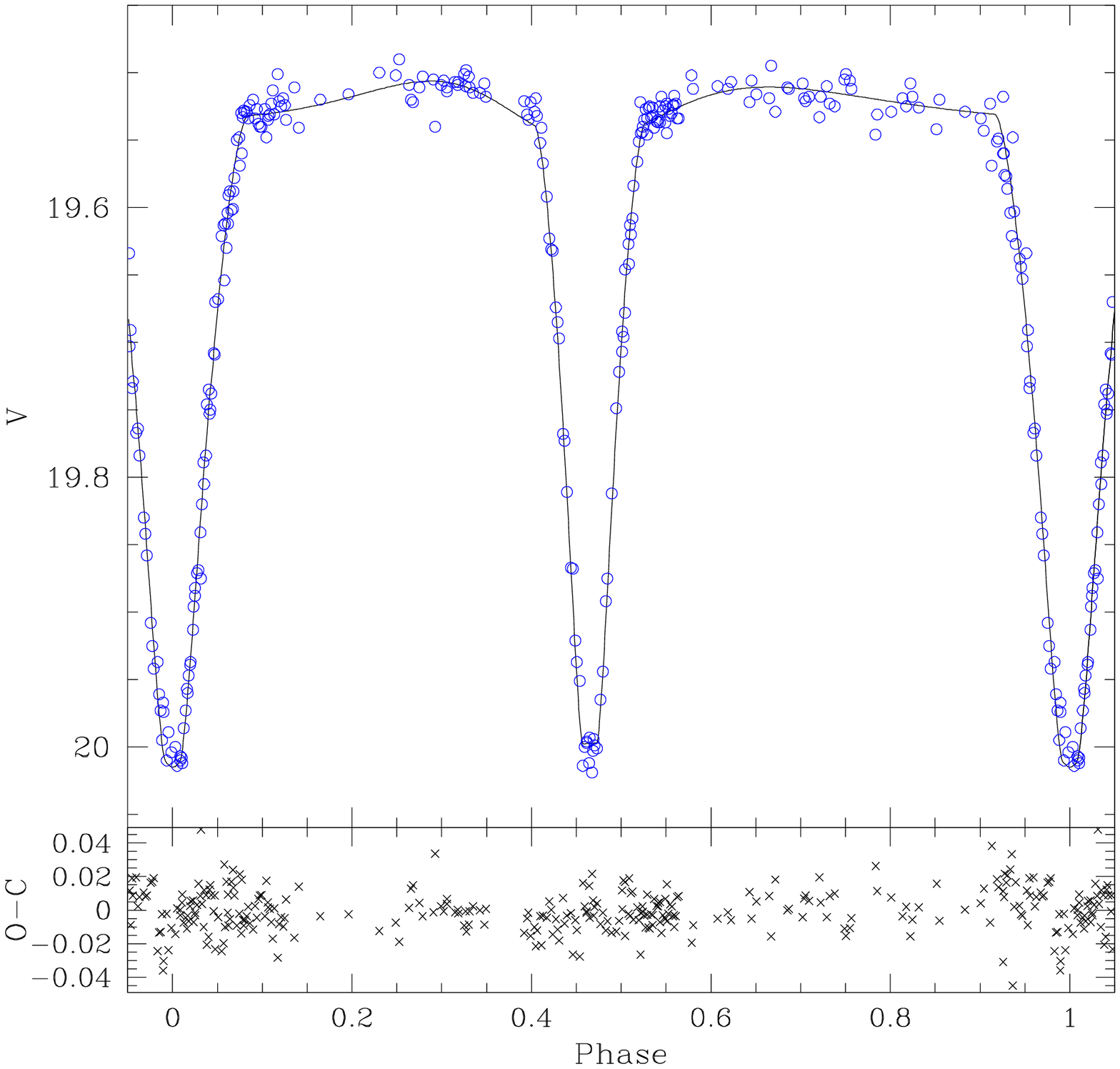}
\caption{$V-$band light curve of the DEB with model fit from the
Wilson-Devinney program. The rms is 0.01 mag.}
\label{lcv}
\end{figure}

\begin{figure}[ht]  
\plotone{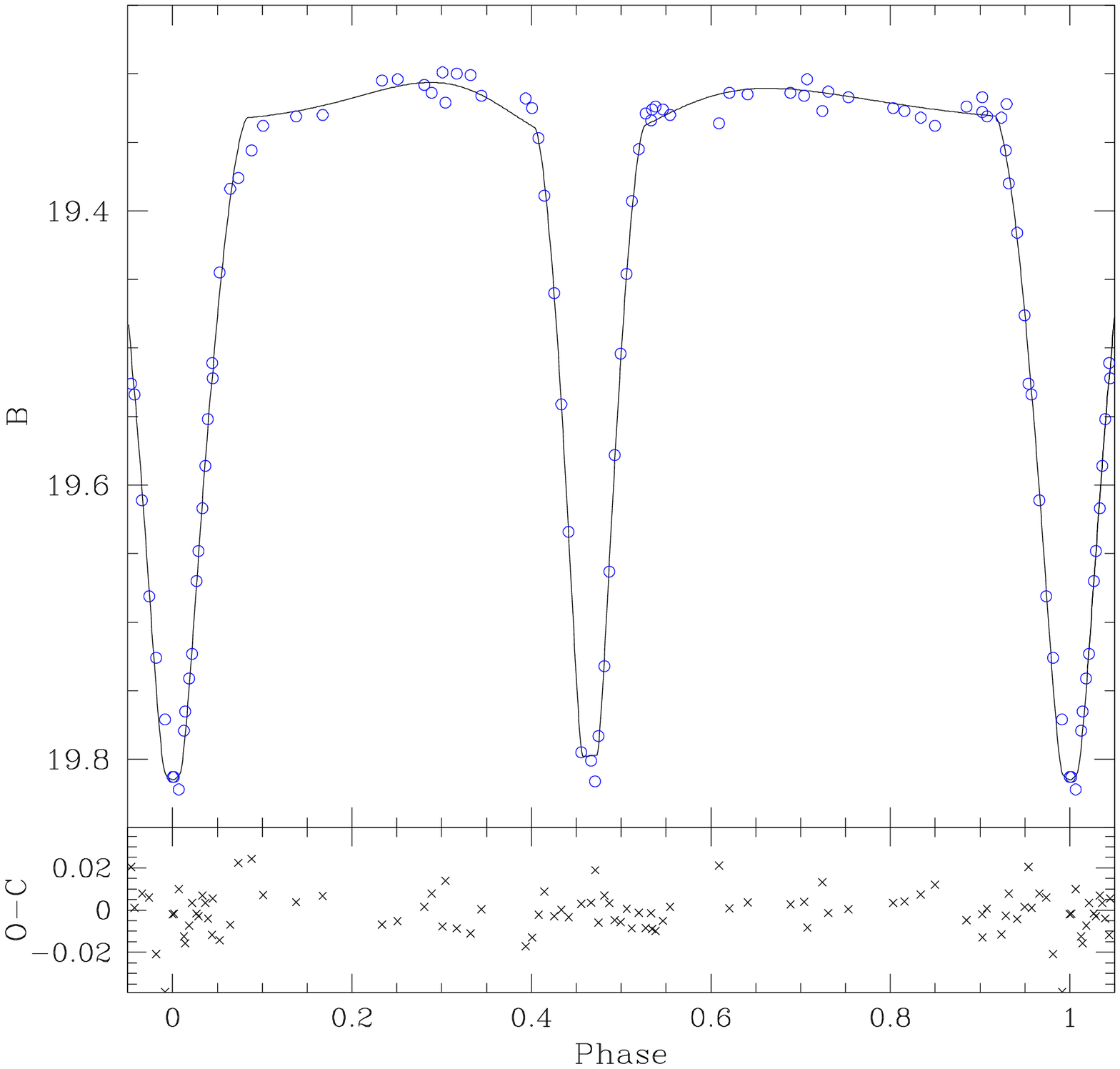}
\caption{$B-$band light curve of the DEB with model fit from the
Wilson-Devinney program. The rms is 0.01 mag.}
\label{lcb}
\end{figure}

\begin{figure}[ht]  
\plotone{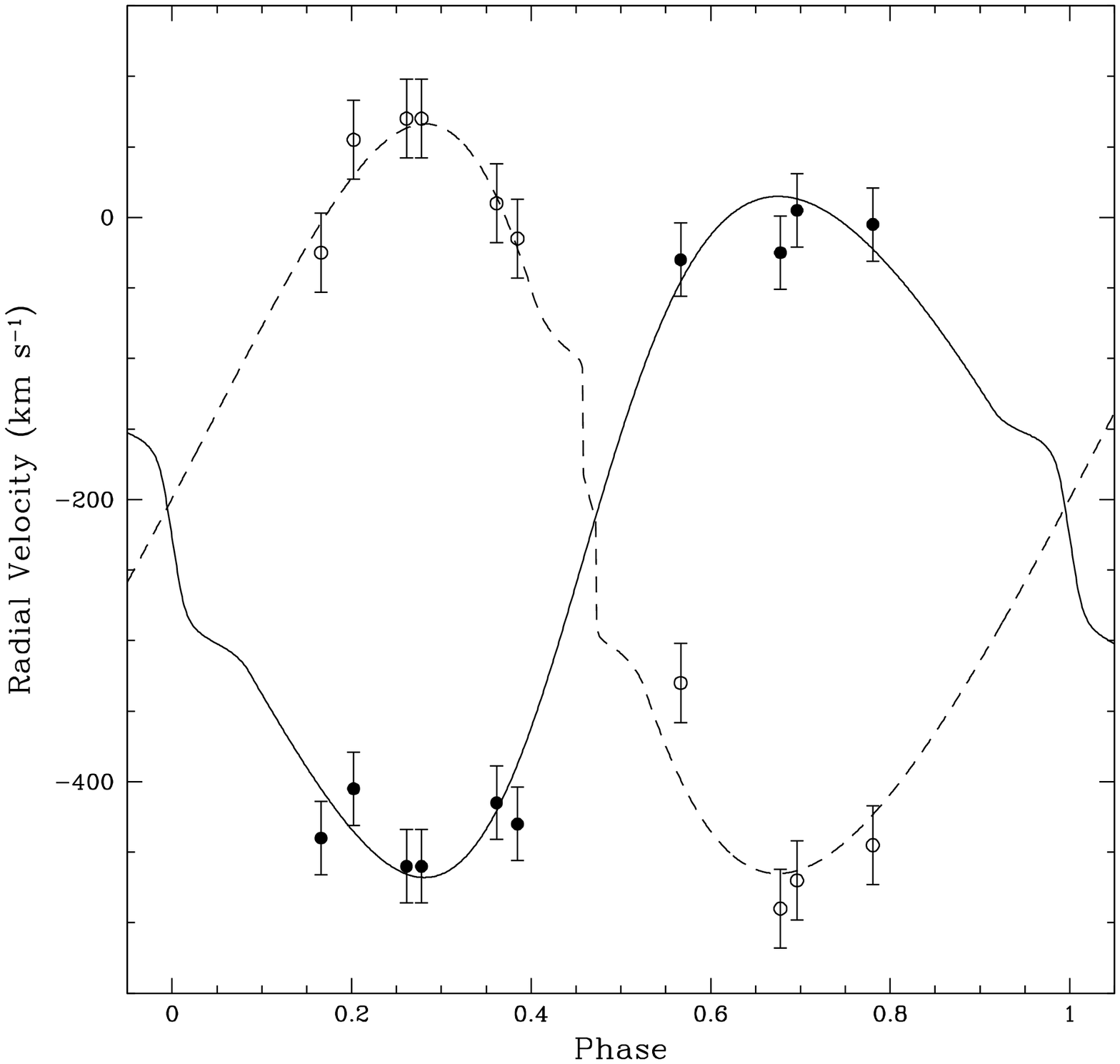}
\caption{Radial velocities for the DEB measured by TODCOR with FASTWIND
synthetic spectra. Model fit is from Wilson-Devinney program, including
the Rossiter effect. Error bars correspond to the rms of the fit: 26.0
$\rm km\; s^{-1}$ for the primary (filled circles) and 28.0 $\rm km\;
s^{-1}$ for the secondary (open circles).}
\label{rv}
\end{figure}

We defined convergence to be reached after three consecutive iterations
for which the corrections for all adjusted parameters were smaller than
their respective standard or statistical errors
\citep[e.g.,][]{Kallrath99}. The results of the fit and 1 $\sigma$
errors are given in Table~\ref{bv} and the derived physical parameters
in Table~\ref{derived}. The rms residuals were 0.01 for both B and V
light curves. Figures~\ref{lcv} and \ref{lcb} show the $V$ and $B-$band
light curve model fits for the DEB. Note that the deviation of the
secondary eclipse from phase 0.5 is due to the eccentricity of the
system. The radial velocity curve is presented in Figure~\ref{rv}. The
rms residuals are 26.0 $\rm km\; s^{-1}$ for the primary and 28.0 $\rm
km\; s^{-1}$ for the secondary star. The ephemeris is:
\begin{equation}
\rm T(HJD)=2451451.4040(5) + 4.89380(3) \times E,
\end{equation}

The stars are nearly spherical; the radius of the primary varies by
$2\%$ depending on the direction (towards the pole, inner Lagrangian
point, side, back), while the variation in the secondary is less than
$1\%$. We used the volume radius to compute the surface area of the
stars for our distance calculation. The WD values yield masses $\rm
M_{1}=33.4\pm3.5 \; \msun$, $\rm M_{2}=30.0\pm3.3\; \msun$, radii $\rm
R_{1}=12.3\pm0.4\; \rsun$, $\rm R_{2}=8.8\pm0.3\; \rsun$, and values
for $\log(g)$ of 3.78 for the primary and 4.03 for the secondary. The
flux ratio derived from the luminosity and radius ratio is
$F_{1}/F_{2}=1.040$ in $V$-band and $F_{1}/F_{2}=1.047$ in
$B$-band. This implies a temperature difference of 1000 K to 1500 K in
the temperature range considered and at the gravities of the two
objects.

We initially allowed the third light to be a free parameter, however,
after converging to negative values, we set it to zero. The presence
of deep eclipses additionally gives us confidence that there is no
significant blending/crowding effect present. We fixed $d\omega/dt$ to
zero since the residuals did not reveal any significant trend from the
two year baseline of our photometry.

\subsection{Determination of Effective Temperatures}
\label{fastwind}

The effective temperatures of the stars were determined by fitting the
observed spectra with FASTWIND spectra. We adopted solar metallicity in
the models appropriate for a deprojected galactocentric radius of
$460\arcsec$ for the DEB, according to \citet{Zaritsky89,
Urbaneja05}. The value of $\log(g)$ and the flux ratio were both fixed
to the values found from the analysis of the light and radial velocity
curves (see \S\ref{combwd}). We used $v_{turb}$ = 0 as appropriate
\citep[see][]{Repolust04} and $N(He)/N(H) = 0.1$, the measured $v \;
\sin i$ of $120\; \rm km \; s^{-1}$, and adopted values for the mass
loss rates of $\rm \dot{M} = 6.7 \cdot 10^{-7}\; \msun\; yr^{-1}$ for
the primary and $\rm 1.7 \cdot 10^{-7}\; \msun\; yr^{-1}$ for the cooler
secondary. This choice of mass-loss rates is justified by the fit of the
mass-loss dependent HeII line $\lambda$4686 \citep[see][]{Repolust05,
Massey04, Massey05}, shown in Figure~\ref{HeII068}. It also agrees with
the average values found by \citet{Repolust05} for stars of similar
parameters. The terminal velocities of the stellar wind are correlated
with the photospheric escape velocities. Following \citet{Kudritzki00},
we adopted for the former 2200 km$/$s and 2600 km$/$s.

We combined spectra near quadratures (phases 0.26 and 0.68) with similar
radial velocities, degrading the ESI spectra to match GMOS resolution
and dispersion where appropriate. The difference in the velocities is
between $0-60\; \rm km \; s^{-1}$, which corresponds to a maximum shift
of $0.9\; \rm \AA$ in our wavelength range, and does not affect the
temperature determination. We fit composite model spectra to all the
helium lines in our wavelength range except HeI $\lambda$4713, which was
too noisy to use quantitatively. However, the fits were in agreement
with this line as well. The composite spectra were calculated as
follows:

\begin{equation}
\rm F_{\lambda} = w_{1}\; F_{1,\lambda}(T_{1}) + w_{2}\;
F_{2,\lambda}(T_{2})
\end{equation}
where
\begin{equation}
\rm w_{1} = \frac{1}{1 + (R_{1}/R_{2})^2\; F_{2}/F_{1}}
\;\; and \;\; w_{2} = 1 - w_{1}
\end{equation}

\begin{figure}[ht]  
\includegraphics[angle=90, width=7in]{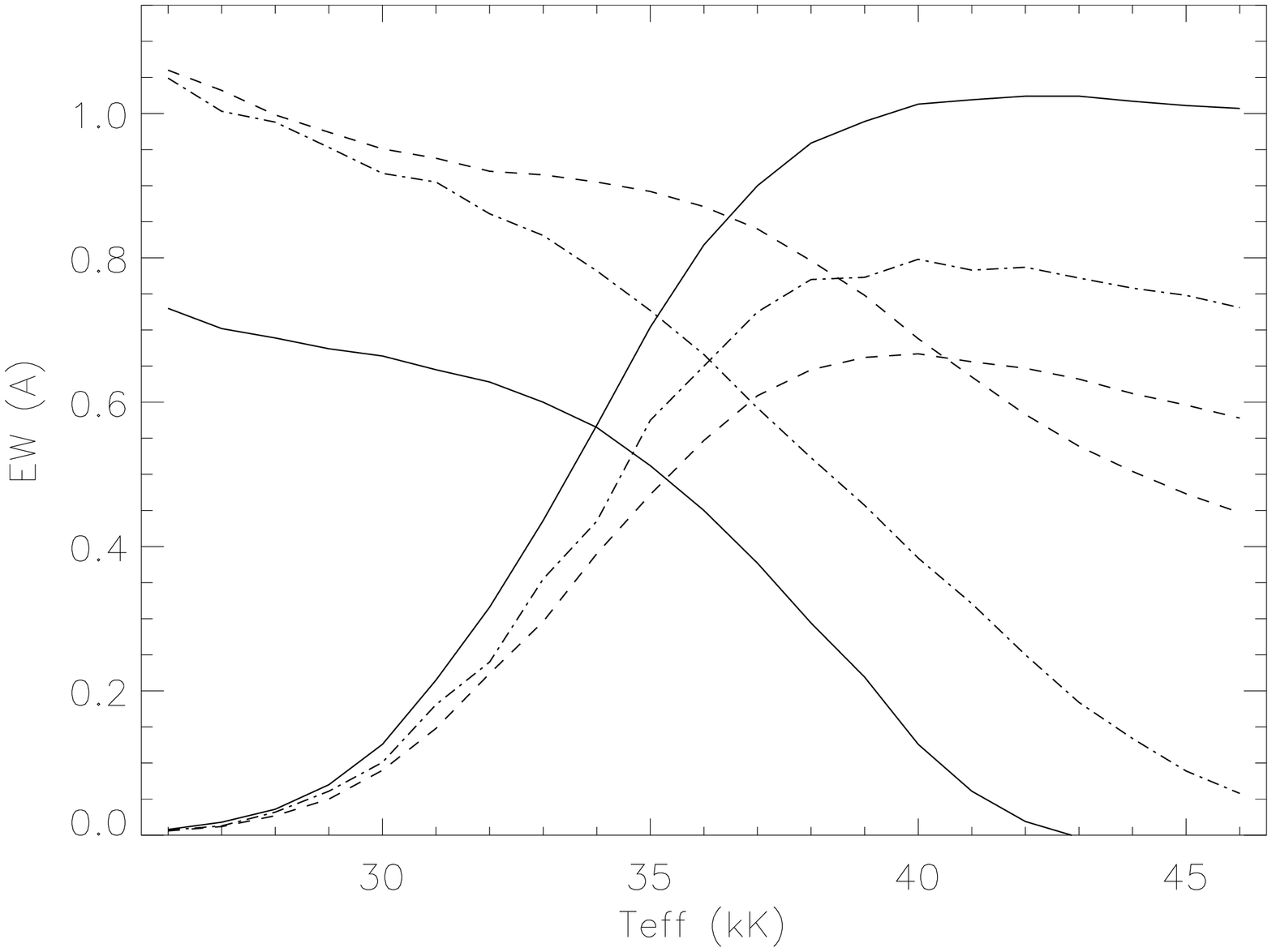}
\caption{Equivalent widths of helium lines vs. $\rm T_{eff}$. The HeI
line strengths decrease with increasing $\rm T_{eff}$ (solid:
$\lambda$4921, dashed-dotted: $\lambda$4471, dashed: $\lambda$4026); the
HeII line strengths increase (solid: $\lambda$5411, dashed-dotted:
$\lambda$4542, dashed: $\lambda$4200).}
\label{eqwidth}
\end{figure}

The ratio of the radii and fluxes were fixed to the values derived in
the previous Section, therefore the only free parameter was T$_1$. The
HeII $\lambda4542/$HeI $\lambda4471$ ratio measured from the spectra
indicated a spectral type of O7 for both stars. In this regime, the
helium line strengths are quite sensitive to temperature. The HeII line
equivalent widths increase with increasing effective temperature,
whereas the HeI widths decrease. Figure~\ref{eqwidth} illustrates this
dependence for several lines. Finally, we resolved the degeneracy in the
radii from the information in the spectra by determining that the
slightly hotter primary contributes twice as much light, thus forcing it
to be the larger star.

\begin{figure}[ht]  
\includegraphics[angle=90,width=7in]{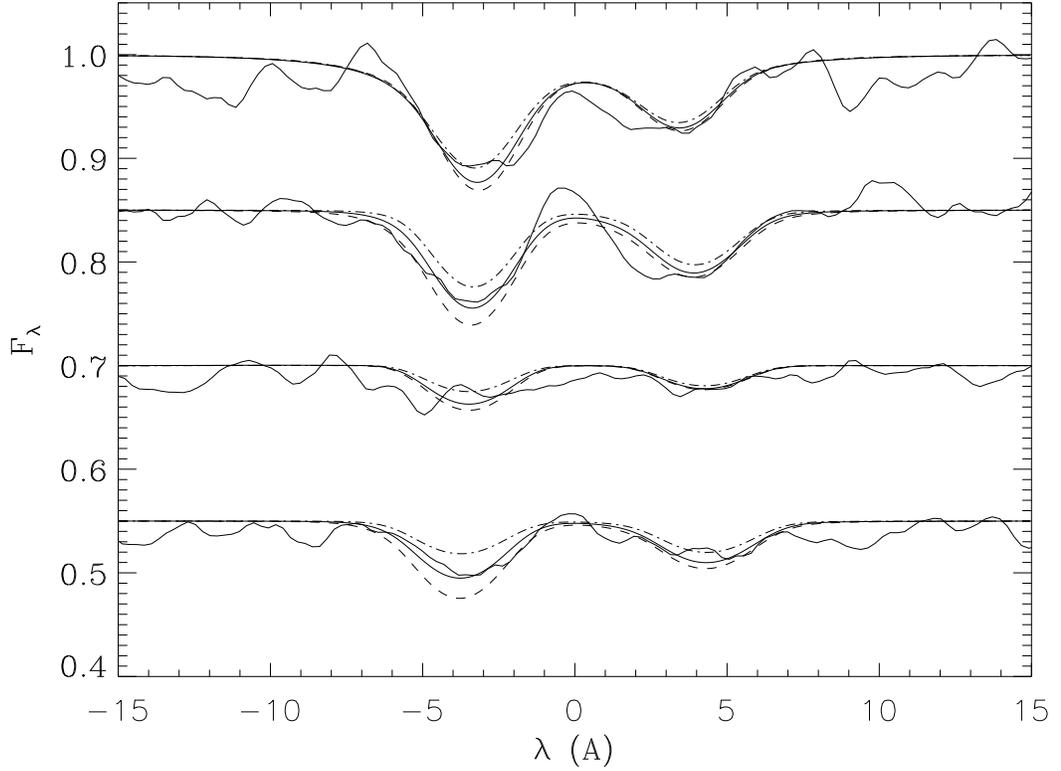}
\caption{Fit of the composite spectra of HeI lines at phase 0.26. At
this phase the absorption lines originating in the primary are
blue-shifted, whereas the lines from the secondary are red-shifted.
Rectified spectra corrected for the gamma velocity are plotted as a
function of wavelength displacement from the atomic line transition
wavelength. The lines from top to bottom are HeI $\lambda\lambda$4026,
4471, 4713, 4921 and have been shifted in flux by 0.15 for clarity.
Overplotted are the model calculations as described in the text. The
solid curve represents our final model corresponding to a primary
effective temperature of 37000K. The dashed and the dashed-dotted curves
represent primary temperatures of 39000K and 35000K, respectively.}
\label{HeI025}
\end{figure}

\begin{figure}[ht]  
\includegraphics[angle=90,width=7in]{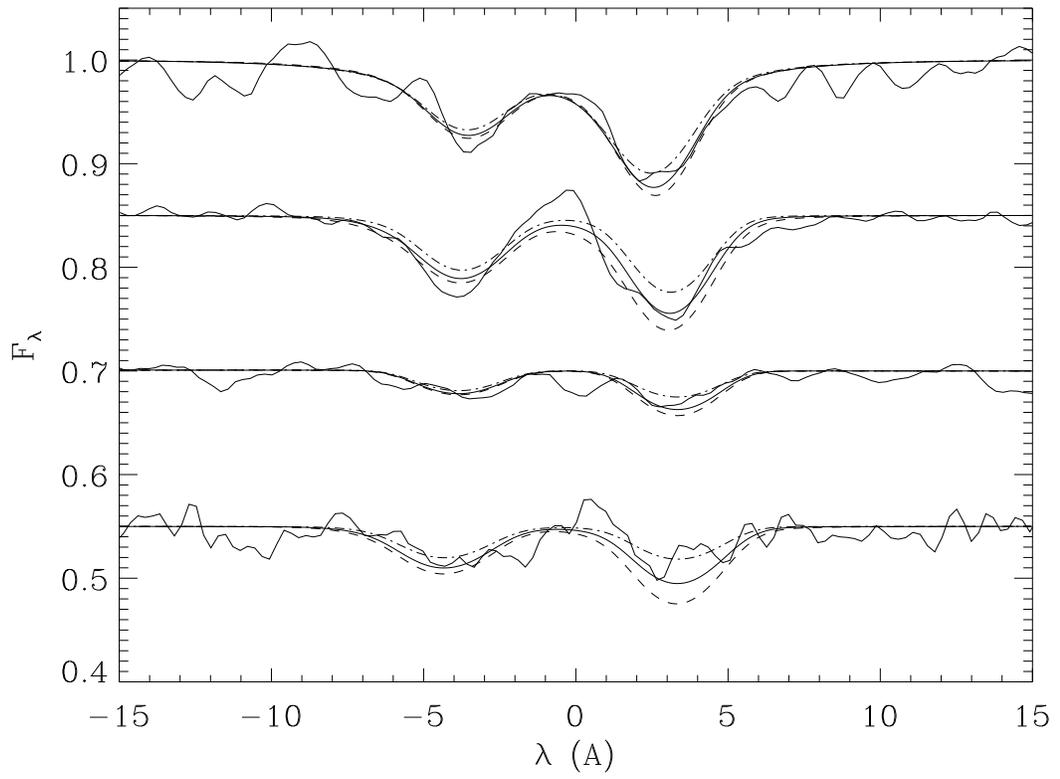}
\caption{Same as Figure~\ref{HeI025}, but at phase 0.68, where the lines
from the primary (secondary) are red-shifted (blue-shifted).}
\label{HeI068}
\end{figure}

\begin{figure}[ht]  
\includegraphics[angle=90,width=7in]{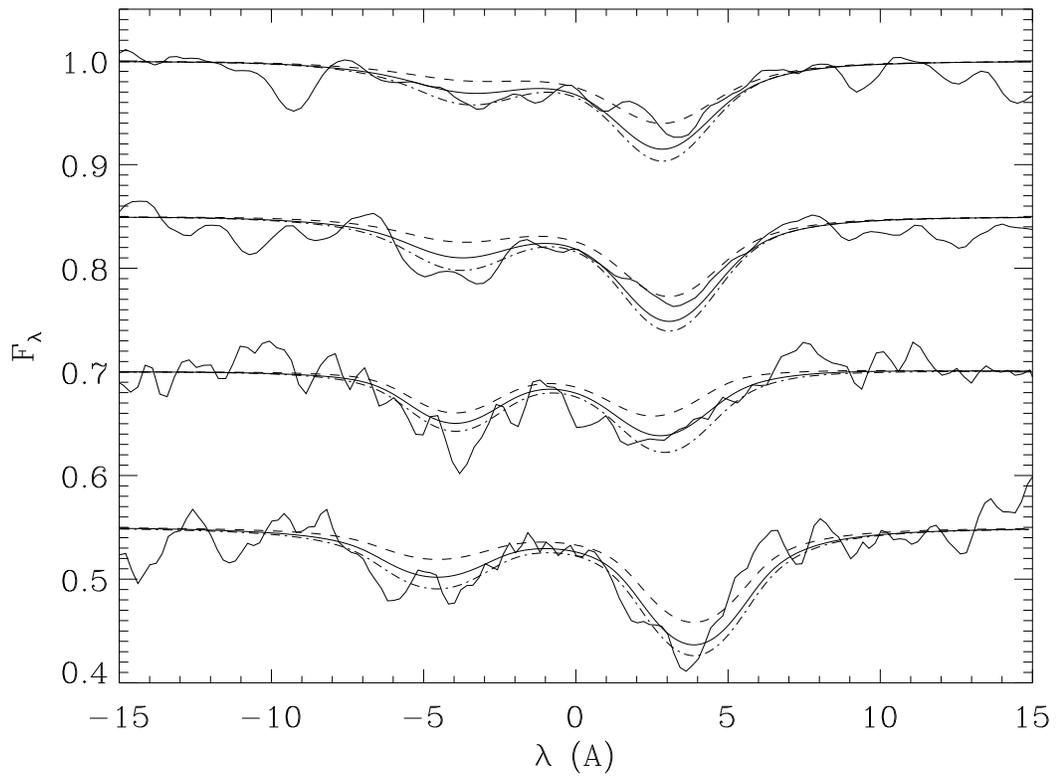}
\caption{Same as Figure~\ref{HeI068}, but for the HeII lines. Plotted are
HeII $\lambda\lambda$4200, 4542, 4686, 5411 from top to bottom.}
\label{HeII068}
\end{figure}

\begin{figure}[ht]  
\includegraphics[angle=90,width=7in]{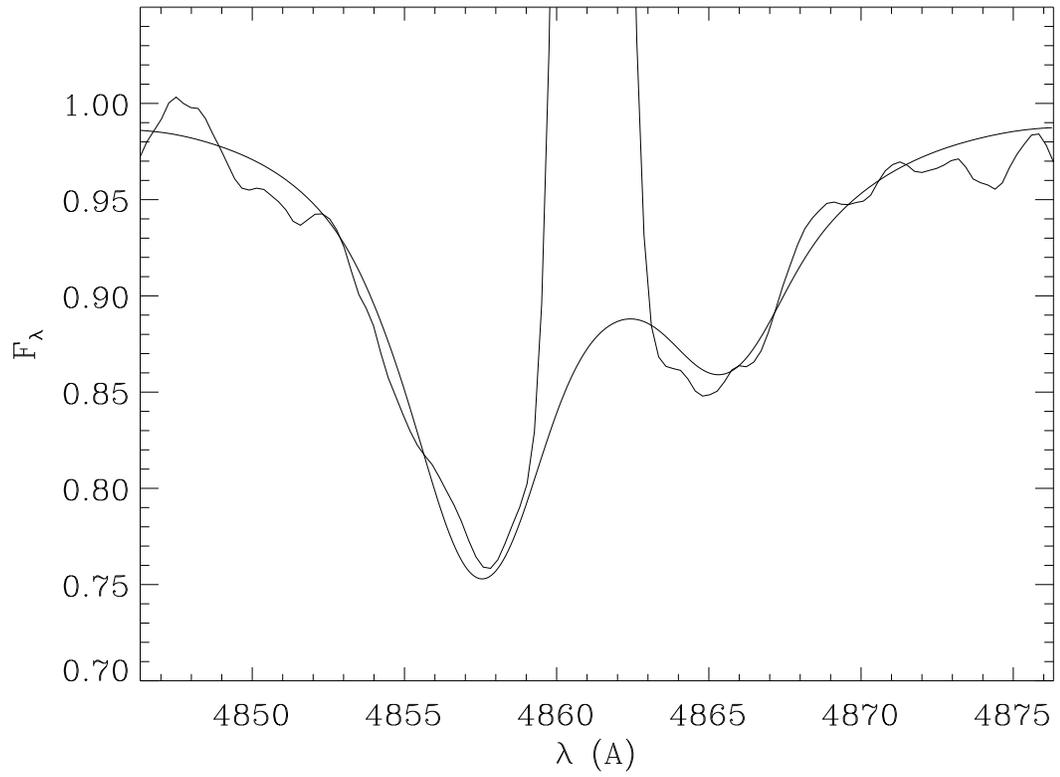}
\caption{Plot of the H$\beta$ profile at phase 0.26 compared to the final model.}
\label{Hbeta025}
\end{figure}

\begin{figure}[ht]  
\includegraphics[angle=90,width=7in]{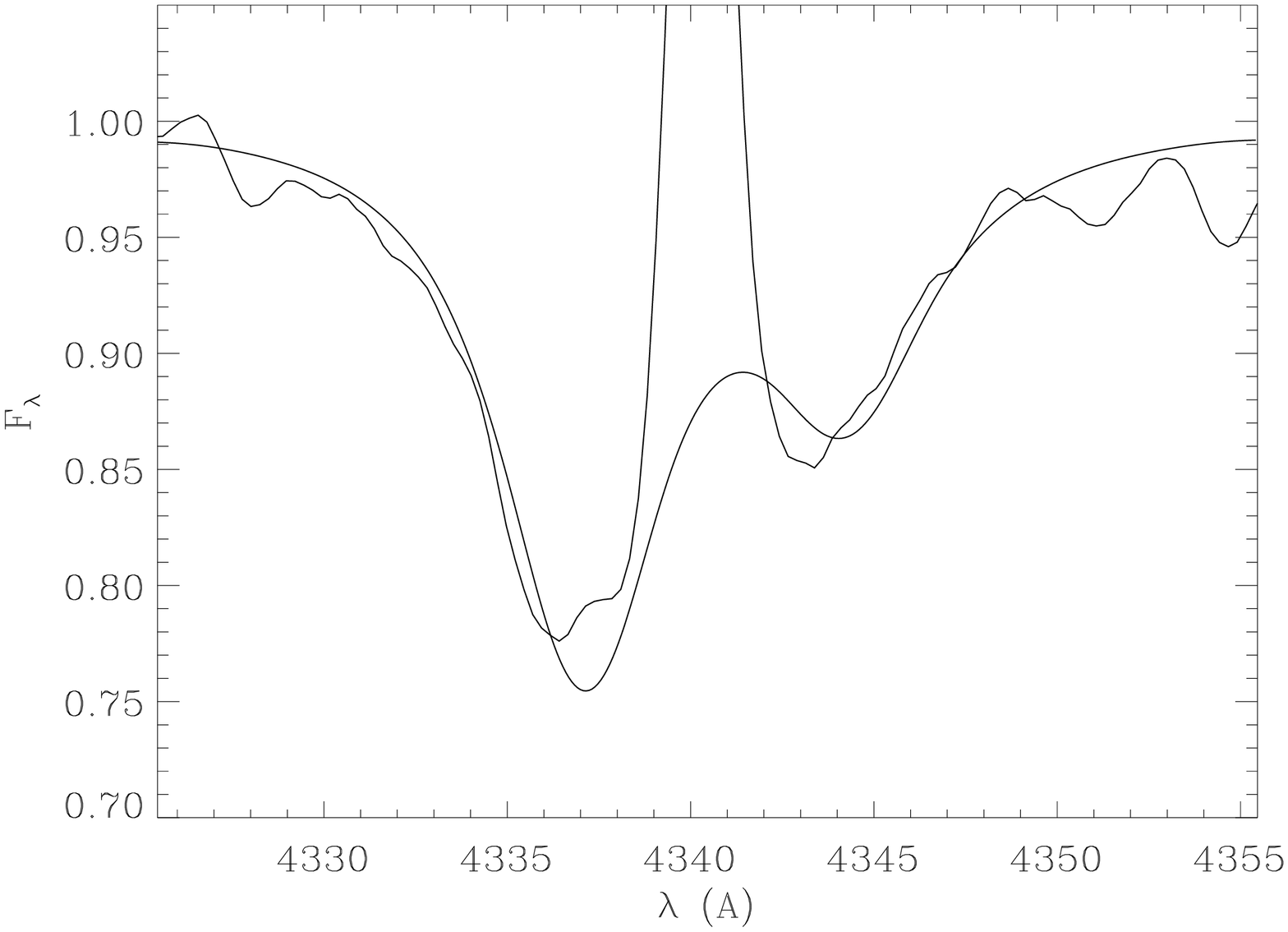}
\caption{Same as Figure~\ref{Hbeta025}, but for H$\gamma$.}
\label{Hgamma025}
\end{figure}

The temperature of the primary was derived from a simultaneous fit to
the following 7 lines: HeI $\lambda\lambda$4026, 4388, 4471, 4921 and
HeII $\lambda\lambda$4200, 4542, 4686. Figures~\ref{HeI025},
\ref{HeI068} and \ref{HeII068} show fits to the helium lines at each
quadrature. The fits to the Balmer lines shown in Figures~\ref{Hbeta025}
and \ref{Hgamma025} demonstrate that the model atmospheres reproduce the
spectra correctly with the gravities derived from the light and radial
velocity curve analysis. The strong emission lines in the center of the
hydrogen and some of the helium lines originate in nearby \ion{H}{2}
regions. The effective temperature of the secondary follows from the
flux ratio and is consistent with the spectral fit of the lines from the
secondary as displayed in the Figures.  We found the best fit
temperature for all the helium lines to be $T_{\rm eff,1} =
37000\pm1500$ K, and consequently $T_{\rm eff,2} = 35600\pm1500$~K.

\section{Distance Determination}

The flux $f_{\lambda}$ measured at Earth at a certain wavelength
$\lambda$ from a binary at distance $d$ is given by

\begin{equation}
\label{disteq}
f_{\lambda}=\frac{1}{d^2} \left(R_{1}^2 \; F_{1,\lambda}+R_{2}^2 \;
F_{2,\lambda} \right) \times 10^{-0.4 \; A\left(\lambda \right)},
\end{equation} 

\noindent where $R_{1}$ and $R_{2}$ are the radii of the two stars and
$F_{1,\lambda}$ and $F_{2,\lambda}$ the surface fluxes. The total
extinction $A\left(\lambda \right)$ is a function of the reddening
$E(B-V)$, the normalized extinction curve $k(\lambda-V) \equiv
E(\lambda-V)/E(B-V)$ and the ratio of total to selective extinction in
the $V$ band $R \equiv A(V)/E(B-V)$:

\begin{equation}
A\left(\lambda \right) =
E\left(B-V\right)\left[k\left(\lambda-V\right)+R\right].
\end{equation} 

\begin{figure}[ht]  
\includegraphics[angle=90, width=7in]{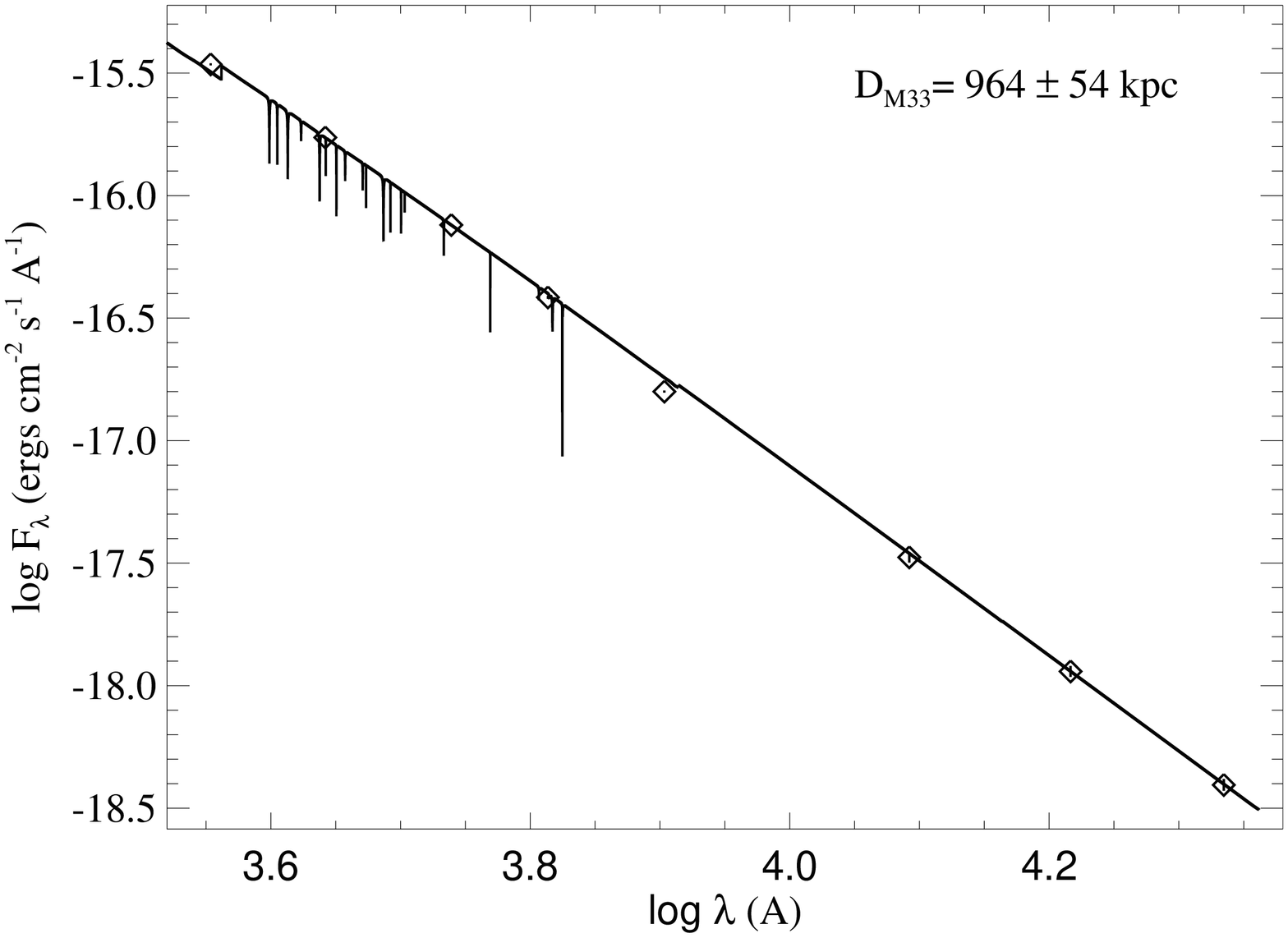}
\caption{Fit of the reddened DEB model spectrum to the $BVRJHK_s$
ground-based photometry. Overplotted is the $U$ and $I$ photometry from
\citet{Massey06}. The best fit values of $E(B-V)=0.09\pm0.01$ and
$R_{V}=3.5\pm0.5$ yield a distance modulus to the DEB and thus M33 of
$24.92\pm0.12$ mag ($964\pm54$ kpc).}
\label{sed}
\end{figure}

\begin{figure}[ht]  
\includegraphics[angle=90, width=7in]{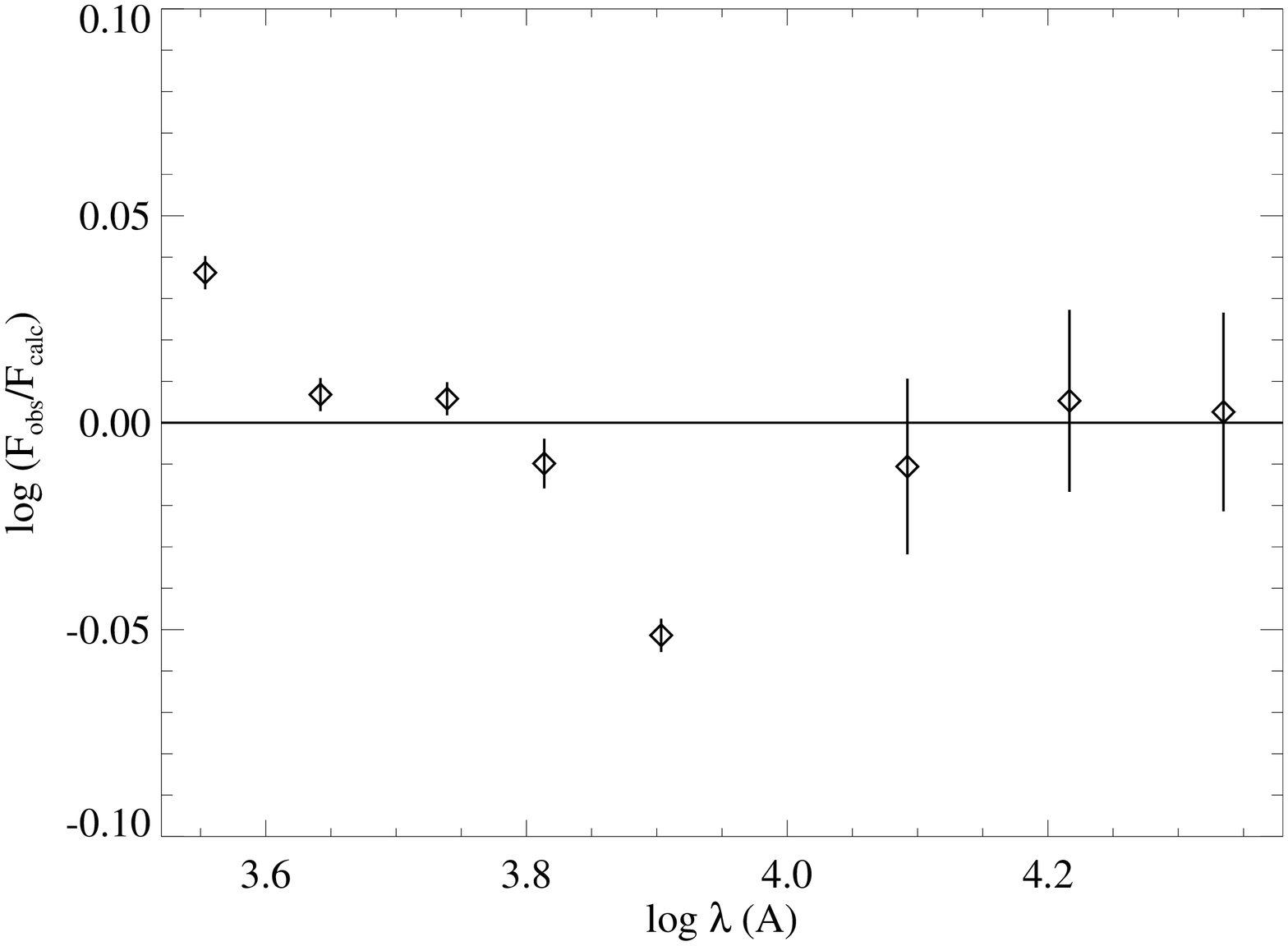}
\caption{Residuals of the SED fit, in terms of the flux ratio. The error
in the fit to the $BVRJHK_s$ photometry corresponds to 0.01 mag. $U$ and
$I$ are shown, but not used in the fit. Error bars correspond to the
photometric error for each band in flux units.}
\label{resid}
\end{figure}

Having measured the temperatures of the stars from the spectra, we
computed fluxes and fit to the observed magnitudes, using equation
\ref{disteq} and FASTWIND model atmospheres with $T_{\rm eff,1} = 37000$
K and $T_{\rm eff,2} = 35000$ K. We calculated synthetic photometry of
the composite spectrum over the appropriate Johnson-Cousins optical
filter functions as defined by \citet{Bessell90} and calibrated by
\citet{Landolt92}, and the 2MASS filter set. Monochromatic fluxes were
measured at the isophotal wavelengths \citep[see][]{Tokunaga05}, which
best represent the flux in a passband. We used zeropoints from
\citet[][Appendix A]{Bessell98} and \citet{Cohen03} to convert the
fluxes to magnitudes. We reddened the model spectrum using the reddening
law parameterization of \citet{Cardelli89}, as prescribed in
\citet{Schlegel98}, and simultaneously fit the optical and near-infrared
$BVRJHKs$ photometry. Specifically, we computed the intrinsic
$(B-V)_0=-0.29$ from the model atmospheres at the isophotal wavelengths,
thus yielding $E(B-V)=0.09\pm0.01$. The best fit that minimized the
photometric error over the 6 photometric bands was given by
$R_V=3.5\pm0.5$. The resulting distance modulus to the DEB and thus M33
is $24.92\pm0.12$ mag ($964\pm54$ kpc). The fit of the reddened model
spectrum to the photometry is shown in Figure~\ref{sed} and the
residuals in Figure~\ref{resid}. The $U$ and $I$ values from
\citet{Massey06} are shown but not used in the fit because there are
inherent problems with $U-$band photometry \citep[see discussion
in][]{Massey02b} and the $I-$band often suffers from fringing
effects. The $BVR$ photometry alone, assuming $R_{V}=3.1$ and
$E(B-V)=0.09$, yields a distance modulus of 24.95 mag, demonstrating the
consistency of the near-infrared with the optical photometry.

The uncertainty in the distance was computed by adding in quadrature the
individual conservative errors: $4\%$ in the radii which translates to
0.085 mag in the distance modulus, $4\%$ in $T_{\rm eff,1}$ which
corresponds to 0.06 mag, 0.04 mag from the SED fit, assuming
$R_{V}=3.5\pm0.5$ and $E(B-V)=0.09\pm0.01$, and 0.03 mag in the
flux. The error in the flux results from adding in quadrature the
statistical 0.01 mag uncertainty from the fit to $BVRJHK_s$ and the 0.03
mag uncertainty in the zeropoints. The total uncertainty is thus 0.12
mag. We did not attempt a full statistical treatment, since the error is
dominated by the uncertainty in the radii. Note, that at these high
temperatures the $B-V$ color of stars saturates, giving a weak
dependence of the distance modulus on $\rm T_{eff}$.

As an independent check on our reddening and extinction determination we
used Chorizos \citep{Maiz04}. We first generated a grid of TLUSTY models
\citep{Lanz03} at solar metallicity in the temperature and gravity
regime of interest and fixed the effective temperature to $37000$ K and
$\log(g)=3.80$. The $\chi^2$ minimization fit to the $BVJHK_s$
photometry resulted in $R_{5495}=3.7\pm0.5$ and
$E(4405-5495)=0.07\pm0.01$, in agreement with the FASTWIND
analysis. Note, that we computed these values at the isophotal
wavelengths, which are bluer. We adopted these errors in the error
analysis above.

\section{Discussion}

We present the first distance to a detached eclipsing binary in M33,
establishing it as an independent rung on the cosmological distance
ladder. This distance determination is a significant step towards
replacing the current anchor galaxy of the extragalactic distance scale,
the LMC, with galaxies more similar to those in the HST Key Project
\citep{Freedman01}, such as M33 and M31. We have chosen a detached
eclipsing binary to simplify the modeling and derived a distance modulus
of $24.92\pm0.12$ mag.

D33J013346.2+304439.9 is located in the rich OB~66 association
\citep{Humphreys80}, which contains a relatively high massive star
population \citep{Massey95}. The presence of one of the best candidate
detached eclipsing binaries for distance determination in this
association is not surprising. In addition to the DEB, OB~66 contains
several other eclipsing binaries (see Paper VI), which suggests a high
binary star formation rate for massive stars. Our adopted color excess
value $E(B-V)=0.09\pm0.01$ is smaller than estimations from
\citet{Massey95} for OB~66. Using the ``$q$ method'' and $UBV$
photometry of 36 stars, they derive $E(B-V)=0.15\pm0.02$ and their
spectroscopic sample yields $E(B-V)=0.13\pm0.01$. However, our
multi-band photometry combined with the spectroscopy determines the
reddening accurately, thus it is unlikely our distance estimation
suffers from systematic errors due to reddening.

There are several avenues for improving the distance to M33 and M31
using eclipsing binaries. \citet{Wyithe02} propose the use of
semi-detached eclipsing binaries to be just as good or better distance
indicators as detached eclipsing binaries, which have been traditionally
considered to be ideal. The use of new improved stellar atmosphere
models to derive surface brightnesses versus calibrations based on
interferometry removes the restriction to DEBs for distance
determination. Additionally, \citet{Wyithe02} outline other benefits for
using semi-detached binaries: their orbits are tidally circularized and
their Roche lobe filling configurations provide an extra constraint in
the parameter space, especially for complete eclipses ($i\sim90$
deg). Bright semi-detached binaries in M33 or M31 are not as rare as
DEBs, and are easier to follow-up spectroscopically, as demonstrated by
\citet{Ribas05} in M31. Thus, for the determination of the distances to
M33 and M31 to better than $5\%$ we suggest both determining distances
to other bright DEBs and to semi-detached systems found by DIRECT and
other variability surveys. Additional spectroscopy of the DEB would also
improve the current distance determination to M33, since the errors are
dominated by the uncertainty in the radius or velocity semi-amplitude.

How does our M33 distance compare to previous determinations?
Table~\ref{distances} presents a compilation of 13 recent distance
determinations to M33 ranging from 24.32 to 24.92 mag, including the
reddening values used. Our measurement although completely independent
yields the largest distance with a small 6\% error, thus is not
consistent with some of the previous determinations. This possibly
indicates unaccounted sources of systematic error in the calibration of
certain distance indicators.

The implications of our result on the extragalactic distance scale are
significant, especially when comparing to the $HST$ Key Project
\citep{Freedman01} distance to M33. They derive a metallicity corrected
Cepheid distance of $24.62\pm0.15$ mag, using a high reddening value of
$E(V-I)=0.27$ and an assumed LMC distance modulus of $18.50\pm0.10$ mag.
If we calculate the LMC distance our result would imply, we derive
$18.80\pm0.16$ mag, which is not consistent with the eclipsing binary
determinations. The error is obtained by adding in quadrature the
individual errors in the two distance measurements. Taking this one step
further, our LMC distance would imply a 15\% decrease in the Hubble
constant to $H_{0}=61\; \rm km\;s^{-1}\; Mpc^{-1}$. This improbable
result brings into question the Key Project metallicity corrections and
reddening values not only for M33, but also for the other galaxies in
the Key Project. We thus demonstrate the importance of accurately
calibrating the distance scale and determining $H_{0}$, which are both
vital for constraining the dark energy equation of state \citep{Hu05}
and complementing the cosmic microwave background measurements from the
Wilkinson Microwave Anisotropy Probe \citep[WMAP;][]{Spergel06}.

\acknowledgments{We are very grateful to Bohdan Paczy\'nski for
motivating us to undertake this project almost a decade ago. We thank
Phil Massey for sending us his photometry of the DEB from his Local
Group Survey before publication. We would like to thank the staff of the
Mt. Hopkins observatory for their support during extensive observing
campaigns and the anonymous referee for comments that improved the
manuscript. AZB and KZS were partially supported by HST Grant
HST-GO-09810.06-A. AZB acknowledges research and travel support from the
Carnegie Institution of Washington through a Vera Rubin
Fellowship. Support for L.M.M. was provided by NASA through Hubble
Fellowship grant HST-HF-01153 from the Space Telescope Science Institute
and by the National Science Foundation through a Goldberg Fellowship
from the National Optical Astronomy Observatory. JK was supported by
MNiI grant 1P03D001-28. BJM was supported by the Polish KBN grant
1P03D01230. GT acknowledges partial support from NSF grant AST-0406183.

Some of the data presented herein were obtained at the W.M. Keck
Observatory, which is operated as a scientific partnership among the
California Institute of Technology, the University of California and the
National Aeronautics and Space Administration. The Observatory was made
possible by the generous financial support of the W.M. Keck Foundation.
The authors wish to recognize and acknowledge the very significant
cultural role and reverence that the summit of Mauna Kea has always had
within the indigenous Hawaiian community.  We are most fortunate to have
the opportunity to conduct observations from this mountain. This paper
is also based on observations obtained under Program ID GN-2004B-Q-25
and GN-2005B-DD-4 at the Gemini Observatory, which is operated by the
Association of Universities for Research in Astronomy, Inc., under a
cooperative agreement with the NSF on behalf of the Gemini partnership:
the National Science Foundation (United States), the Particle Physics
and Astronomy Research Council (United Kingdom), the National Research
Council (Canada), CONICYT (Chile), the Australian Research Council
(Australia), CNPq (Brazil) and CONICET (Argentina). This publication
makes use of data products from the Two Micron All Sky Survey, which is
a joint project of the University of Massachusetts and the Infrared
Processing and Analysis Center/California Institute of Technology,
funded by the National Aeronautics and Space Administration and the
National Science Foundation.
}

%%%%%%%%%%%%%%%% BIBLIOGRAPHY  %%%%%%%%%%%%%%%%%%%%%%%%

\clearpage
\input{tab1.tex}
\input{tab2.tex}
\input{tab3.tex}
\input{tab4.tex}
\input{tab5.tex}
\input{tab6.tex}
\input{tab7.tex}
\end{document}

%% file: tab1.tex
\begin{deluxetable}{lcrr}
\tabletypesize{\footnotesize}
\tablewidth{0pc}
\tablecaption{\sc $V-$Band DEB Light Curve}
\tablehead{
\colhead{HJD$^{\rm a}$} & \colhead{$V$} & \colhead{$\sigma_{mag}$} \\}
\startdata
  1451.6862 & 19.654 &  0.007 \\
  1451.6979 & 19.630 &  0.008 \\
  1451.7073 & 19.612 &  0.008 \\
  1451.7241 & 19.602 &  0.008 \\
  1451.7329 & 19.601 &  0.008 \\
\enddata                         
\label{Vlightcurve}
\tablenotetext{a}{HJD-2450000} \tablecomments{Table~\ref{Vlightcurve} is
available in its entirety in the electronic version of the Journal. A
portion is shown here for guidance regarding its form and content.}
\end{deluxetable} 

%% file: tab2.tex
\begin{deluxetable}{lcrr}
\tabletypesize{\footnotesize}
\tablewidth{0pc}
\tablecaption{\sc $B-$Band DEB Light Curve}
\tablehead{
\colhead{HJD$^{\rm a}$} & \colhead{$B$} & \colhead{$\sigma_{mag}$} \\}
\startdata
  1451.7623 &  19.376 &  0.010 \\
  1451.8347 &  19.356 &  0.009 \\
  1451.8981 &  19.338 &  0.007 \\
  1452.8169 &  19.314 &  0.008 \\
  1452.8922 &  19.321 &  0.008 \\
\enddata                         
\label{Blightcurve}
\tablenotetext{a}{HJD-2450000}\tablecomments{Table~\ref{Blightcurve} is
available in its entirety in the electronic version of the Journal. A
portion is shown here for guidance regarding its form and content.}
\end{deluxetable} 

%% file: tab3.tex
\begin{deluxetable}{lllccc}
\tabletypesize{\footnotesize}
\tablewidth{0pc}
\tablecaption{\sc Spectroscopic Observations}
\tablehead{
\colhead{UT Obs.} & \colhead{Telescope} & \colhead{Instr.} &
\colhead{\# exp.} & \colhead{Total exp. time} & \colhead{Mid. Exp.}\\
\colhead{Date} & \colhead{} & \colhead{} & \colhead{} & \colhead{(hours)}
& \colhead{HJD}}
\startdata
20021031 & Keck-2 & ESI    & 4 & 1.53  & 2452578.8611 \\
20021101 & Keck-2 & ESI    & 2 & 0.575 & 2452579.8873 \\
20030903 & Keck-2 & ESI    & 5 & 2.92  & 2452886.0960 \\
20030904 & Keck-2 & ESI    & 4 & 3.25  & 2452887.0532 \\
20030905 & Keck-2 & ESI    & 4 & 3.20  & 2452888.0598 \\
20040816 & Gemini & GMOS-N & 5 & 3.75  & 2453234.0229 \\
20040818 & Gemini & GMOS-N & 5 & 3.75  & 2453236.0637 \\
20040912 & Gemini & GMOS-N & 5 & 3.75  & 2453261.0327 \\
20041009 & Gemini & GMOS-N & 5 & 3.75  & 2453287.9396 \\
20041011 & Keck-2 & ESI    & 7 & 5.25  & 2453289.9871 \\
20041112 & Keck-2 & ESI    & 3 & 4.50  & 2453321.8224 \\
\enddata                         
\label{debtab:obs}
\end{deluxetable} 

%% file: tab4.tex
\begin{deluxetable}{lcrrrr}
\tabletypesize{\footnotesize}
\tablewidth{0pc}
\tablecaption{\sc Heliocentric Radial Velocity Measurements for the DEB}
\tablehead{
\colhead{Mid. Exp.} & \colhead{Orbital Phase} & \colhead{$RV_{1}$} &
\colhead{$RV_{2}$} & \colhead{$(O-C)_1$} & \colhead{$(O-C)_2$} \\
\colhead{HJD} & \colhead{} & \colhead{($km\;s^{-1}$)}&
\colhead{($km\;s^{-1}$)} & \colhead{($km\;s^{-1}$)} & \colhead{($km\;s^{-1}$)}}
\startdata
2452578.8611 & 0.386 & $-430$ & $ -15$  &   $-42$&   $    7$\\
2452579.8873 & 0.595 & $ -20$ & $-560$  &   $ -8$&   $ -124$\\
2452886.0960 & 0.167 & $-440$ & $ -30$  &   $-35$&   $  -21$\\
2452887.0532 & 0.362 & $-415$ & $  10$  &   $  6$&   $   -4$\\
2452888.0598 & 0.568 & $ -35$ & $-320$  &   $ 16$&   $   69$\\
2453234.0229 & 0.260 & $-460$ & $  70$  &   $  5$&   $    7$\\
2453236.0637 & 0.677 & $ -25$ & $-490$  &   $-40$&   $  -25$\\
2453261.0327 & 0.779 & $  -5$ & $-445$  &   $ 17$&   $  -22$\\
2453287.9396 & 0.278 & $-460$ & $  70$  &   $  8$&   $    4$\\
2453289.9871 & 0.698 & $   5$ & $-475$  &   $ -8$&   $   -7$\\
2453321.8224 & 0.203 & $-410$ & $  55$  &   $ 30$&   $   25$\\
\enddata                         
\label{vel}
\end{deluxetable} 

%% file: tab5.tex
\begin{deluxetable}{lc}
\tabletypesize{\footnotesize}
\tablewidth{0pc}
\tablecaption{\sc Results From Combined $LC$ And $RV$ Curve Analysis
With Wilson-Devinney Program}
\tablehead{
\colhead{Parameter} & \colhead{Value}} 
\startdata
Period, P & 4.89380 $\pm$ 0.00003 days \\
Time of primary eclipse, HJD$_0$ & 2451451.4040(5) \\
Inclination, $i$        & 87.2 $\pm$ 0.5 $\deg$\\
Eccentricity, $e$ & 0.18 $\pm$ 0.02 \\
Longitude of periastron, $\omega$ & 252.4 $\pm$ 1.0 deg \\
Surface potential, $\Omega_{1}$ & 5.09 $\pm$ 0.03  \\
Surface potential, $\Omega_{2}$ & 6.29 $\pm$ 0.06 \\
Light ratio in $V$, $L_{2}/L_{1}$ & 0.492 $\pm$  0.005 \\
Light ratio in $B$, $L_{2}/L_{1}$ & 0.493 $\pm$ 0.005 \\
Mass ratio, $q$ & 0.91 $\pm$ 0.07 \\
Systemic velocity, $\gamma$ & $-214\pm7\; \rm km\; s^{-1}$ \\
Semi-major axis, $a$ & $48.4 \pm1.6\; \rsun $  \\
Semi-amplitude, $K_{1}$ & $242\pm11\; \rm km\; s^{-1}$ \\
Semi-amplitude, $K_{2}$ & $266\pm11\; \rm km\; s^{-1}$ \\
Radius, $\rm r_{1,pole}$ & 0.248   $\pm$ 0.002 \\     
............ $\rm r_{1,point}$& 0.267   $\pm$ 0.002  \\
............ $\rm r_{1,side}$ & 0.252   $\pm$ 0.002  \\
............ $\rm r_{1,back}$ & 0.261   $\pm$ 0.002  \\
............ $\rm r_{1}$$^{*}$ &  0.254 $\pm$ 0.002 \\
Radius, $\rm r_{2,pole}$ & 0.180   $\pm$ 0.002  \\     
............ $\rm r_{2,point}$& 0.185   $\pm$ 0.002  \\
............ $\rm r_{2,side}$ & 0.181   $\pm$ 0.002  \\
............ $\rm r_{2,back}$ & 0.184   $\pm$ 0.002  \\
............ $\rm r_{2}$$^{*}$ & 0.182 $\pm$ 0.002 \\
\enddata
\label{bv}
\tablenotetext{*}{Volume radius.}
\end{deluxetable}

%% file: tab6.tex
\begin{deluxetable}{lcc}
\tabletypesize{\footnotesize}
\tablewidth{0pc}
\tablecaption{\sc DEB Physical Parameters}
\tablehead{
\colhead{Parameter} & \colhead{Primary} & \colhead{Secondary}} 
\startdata
Mass ($\msun$) &  $33.4\pm3.5$ & $30.0\pm3.3$ \\
Radius ($\rsun$) & $12.3\pm 0.4$ &  $8.8\pm 0.3$ \\
$\log g$ (cgs) &  $3.78\pm 0.03$ & $4.03\pm 0.03$ \\
$\rm T_{eff}$ (K) & $37000\pm 1500$ & $35600\pm 1500$ \\
$\log L/\lsun$ & $5.41\pm0.17$ & $5.05\pm0.18$ \\ 
\enddata
\label{derived}
\end{deluxetable}

%% file: tab7.tex
\begin{deluxetable}{lccc}
\tabletypesize{\footnotesize}
\tablewidth{0pc}
\tablecaption{\sc Recent Distance Determinations to M33}
\tablehead{
\colhead{Study} & \colhead{Method$^{\rm a}$} & \colhead{Distance Modulus} & \colhead{Reddening}} 
\startdata
This Work & DEB & $24.92\pm0.12$ & $E(B-V)=0.09\pm0.01$ \\
\citet{Sarajedini06} & RR Lyrae & $24.67\pm0.08$ & $\sigma_{E(V-I)}=0.30$\\
\citet{Brunthaler05} & Water Masers & $24.32\pm0.45$ & --- \\
\citet{Ciardullo04} & PNe & $24.86^{+0.07}_{-0.11}$ & $E(B-V)=0.04$ \\
\citet{Galleti04} & TRGB & $24.64\pm0.15$ &$E(B-V)=0.04$  \\
\citet{McConnachie04} & TRGB & $24.50\pm0.06$ &$E(B-V)=0.042$ \\
\citet{Tiede04} & TRGB & $24.69\pm0.07$ &$E(B-V)=0.06\pm0.02$ \\
\citet{Kim02} & TRGB & $24.81\pm0.04(r)^{+0.15}_{-0.11}(s)$ & $E(B-V)=0.04$\\
\citet{Kim02} & RC & $24.80\pm0.04(r)\pm0.05(s)$ & $E(B-V)=0.04$ \\
\citet{Lee02} & Cepheids & $24.52\pm0.14(r)\pm0.13(s)$ &
$E(B-V)=0.20\pm0.04$ \\
\citet{Freedman01} & Cepheids & $24.62\pm0.15$ & $E(V-I)=0.27$ \\
\citet{Pierce00} & LPVs & $24.85\pm0.13$ & $E(B-V)=0.10$ \\
\citet{Sarajedini00} & HB & $24.84\pm0.16$ & $<E(V-I)>=0.06\pm0.02$ \\
\enddata
\label{distances}
\tablenotetext{a}{DEB: detached eclipsing binary; TRGB: tip of the red
giant branch; PNe: planetary nebulae; RC: the red clump; LPVs: long
period variables; HB: horizontal branch stars.}
\end{deluxetable}